\documentclass[12pt]{article}

 \newread\testifexists
 \def\GetIfExists #1 {\immediate\openin\testifexists=#1
     \ifeof\testifexists\immediate\closein\testifexists\else
     \immediate\closein\testifexists\input #1\fi}

 \usepackage{gthstyle}\usepackage{amsfonts}
 \usepackage{amssymb}
 \usepackage{graphicx} \usepackage{epstopdf}
 \def\Bbb#1{\setbox0=\hbox{$\tt #1$}  \copy0\kern-\wd0\kern .1em\copy0}
 \def\bbf#1{\setbox0=\hbox{$#1$} \kern-.025em\copy0\kern-\wd0
         \kern.05em\copy0\kern-\wd0 \kern-.025em\raise.0433em\box0}
 \immediate\openin\testifexists=a
     \ifeof\testifexists\immediate\closein\testifexists\else
     \immediate\closein\testifexists\input a\fimssym.def  % for \Bbb A - Z %%

               \newcommand{\Tr}{{\mbox{Tr}}\,}
                     \newcommand{\fn}{\footnote}
              \newcommand{\nm}{\nonumber}
 \newcommand{\be}{\begin{eqnarray}}             \newcommand{\ee}{\end{eqnarray}}
 \newcommand{\bi}[1]{\begin{itemize}\item[#1]}         \newcommand{\itm}[1]{\item[#1]}  \newcommand{\ei}{\end{itemize}}
 \newcommand{\eqn}[1]{(\ref{#1})}

 \newcommand{\crlb}[1]{\label{#1}\\[2pt]}
 \newcommand{\eela}[1]{\quad\hbox{\scriptsize{#1}}\label{#1}\end{eqnarray}}
 \newcommand{\eelb}[1]{\label{#1}\end{eqnarray}}
 
 \newcommand{\newsecb}[2]{\section{#1}\label{#2}\setcounter{equation}{0}}

 \newcommand{\nolabels} {\def\eel{\eelb} \def\crl{\crlb} \def\newsecl{\newsecb}\def\bibiteml{\bibitem}\def\citel{\cite}}

\newcommand\publishversion{\nolabels\setlength{\textheight}{9in}\setlength{\oddsidemargin}{0in}
    \setlength{\textwidth}{6.3in}\setlength{\topmargin}{-0.1in}}

 \def\a{\alpha}         \def\g{\gamma}      
 \def\d{\delta}         
 \def\k{\kappa}            \def\m{\mu}
 \def\f{\phi}                
 \def\j{\psi}                 \def\s{\sigma}  
 \def\t{\tau}

 \def\HH{{\mathcal H}}    \def\OO{{\mathcal O}}      \def\eps{\epsilon}
 \def\pa{\partial} \def\ra{\rightarrow} 
 \def\bal{$\bullet$} 
 \def\dd{{\rm d}}  \def\bra{\langle}   \def\ket{\rangle}

 \def\qu{\ {\buildrel {\displaystyle ?} \over =}\ }  
 
 \def\iss{\ =\ }

 \def\fract#1#2{{\textstyle{#1\over#2}}}
 \def\ffract#1#2{\raise .2 em\hbox{$\scriptstyle#1$}\kern-.3em/
                 \kern-.2em\lower .15 em \hbox{$\scriptstyle#2$}}
 
 \def\half{\fract12} \def\quart{\fract14} 
 
 \def\part#1#2{{\partial#1\over\partial#2}}

\publishversion
\begin{document} \begin{titlepage}

\title{\normalsize  \hfill ITP-UU-13/22    \\ \hfill SPIN-13/15  %  \\ \hfill arXiv: \(\cdots\)
\vskip 20mm \Large\bf The Fate of the Quantum\fn{Submitted to the Proceedings of the Conference on \textit{Time and Matter}, Venice, March 2013.}}

\author{Gerard 't~Hooft}
\date{\normalsize Institute for Theoretical Physics \\
Utrecht University \\ and
\medskip \\ Spinoza Institute \\ Postbox 80.195 \\ 3508 TD Utrecht, the Netherlands \smallskip \\
e-mail: \tt g.thooft@uu.nl \\ internet: \tt
http://www.phys.uu.nl/\~{}thooft/}

\maketitle

\begin{quotation} \noindent {\large\bf Abstract } \medskip \\

Although the suspicion that quantum mechanics is emergent has been lingering for a long time, only now we begin to understand how a bridge between classical and quantum mechanics might be squared with BellÕs inequalities and other conceptual obstacles. 
Here, it is shown how mappings can be formulated that relate quantum systems to classical systems. By generalizing these ideas, one gets quite general models in which quantum mechanics and classical mechanics can merge.
It is helpful to have some good model examples such as string theory. It is suggested that notions such as `super determinism' and `conspiracy' should be looked at much more carefully than in the, by now, standard arguments.\end{quotation}

\vfill \flushleft August 3, 2013.
\end{titlepage}
\eject

\def\ol{\overline}  			\def\E{\epsilon}		%\def\E{\hbox{\scriptsize{E}}}  	
\def\qp{\mathrm{qp}}  \def\PQ{\mathrm{PQ}}  \def\edge{\mathrm{edge}}
\def\ds{\displaystyle}	\def\low#1{{\raisebox{-3pt}{\scriptsize{$#1$}}}}
\def\o{\_\!\_\!\_\!}   \def\M{\Box\!}   \def\bpm{\begin{pmatrix}} \def\epm{\end{pmatrix}}

%%%%%%%%%%%%%%%%%%%%%%%%%%%%%%%%%%%%%%%%%%%%%%%%%%%%

\newsecl{Introduction. The cellular Automaton Interpretation of \\ Quantum Mechanics}{intro}
	What follows may seem to be a prototype of the most simplistic hidden variable ideas.\cite{GtHCA} Indeed, it is precisely the thing that is being categorically dismissed in the physics community because it seems to lack the conditions for entangled states to arise naturally, and it seems to entail Bell inequalities. As for the first objection, we can bring forward that it is easy to consider entangled states in this formalism, while the problem of Bell's inequalities is more mysterious in this context. Rather than attempting to play down the Bell inequality arguments, we note that our models as yet will be too simple to allow a complete discussion of the relevant gedanken experiments: we cannot do such gedanken experiments because the interactions needed have not yet been introduced. We will observe that, at our level of the discussion, distinctions between `fundamentally' classical and `fundamentally' quantum mechanical models cannot yet be made, and precisely this will make the discussion interesting. The dividing line between classical and quantum mechanical is much more delicate than usually implicated, and such observations may well become helpful and important in the future. In short, we believe this is a way to go. The reader, however, is ask to uphold his/her intuitive objections until the nature of the discussion here is made more clear.
	
The `cellular Automaton (CA) Interpretation'	is then based on some simple axioms:
	\bi{(i)} A cellular automaton describes states \(\vec Q_i\ ,\ i=1,2,\,\cdots\), that evolve entirely classically. Classical data are divided over `cells' that may form a \(D-1\)-dimensional lattice (when excluding the time dimension). For clarity, we take time to be discrete, \(t=t_1\ra t_2\ra \ \cdots\), allowing the classical states to evolve according to some evolution law: \(\vec Q_1\ra\vec Q_2\ra\ \cdots\)\ . Usually, we will require \emph{locality}, which means that the evolution law for each cell only implies nearest neighbors. 
	\itm{(ii)} We  write these states as basis elements of a Hilbert space,   \(|\vec Q_1\ket,\ |\vec Q_2\ket,\)  \(\cdots\), and we allow ourselves to consider `quantum superpositions', \(|\j\ket=\a_1|\vec Q_1\ket+\a_2|\vec Q_2\ket+\cdots\), where the coefficients \(\a_i\) mean nothing more than that \(|\a_i|^2\) stand for the probabilities, while the phases of \(\a_i\) are totally meaningless at this point. These phases only serve to allow us to perform mathematical operations between these states (which are important, as will become manifest shortly).
	\itm{(iii)} We now write the evolution operator \(U\) as a matrix in this Hilbert space:\\  \(|\vec Q(t+\d t)\ket=U(\d t)|\vec Q(t)\ket\ .\) For example, we can order the states \(|\vec Q_i\ket\) such that the matrix \(U\) takes the form
	\be U=\bpm{ 0&0&\cdots&1\cr   1&0&&0\cr 0&1&& 0\cr&&\cdots& } \epm .\eel{Umatrix}
	\itm{(iv)} By inspecting the eigen states of this matrix, it is easy to construct an operator \(H\) such that 
	\(U=\exp(-iH\,\d t)\).
	\itm{(v)} We allow ourselves to perform unitary transformations to \emph{any} other basis. One finds a ``quantum mechanical" system, obeying Schr\"odinger's equation 
	\be {\dd\over\dd t}|\j(t)\ket=-iH|\j(t)\ket\ . \eel{schro}
\ei
These sets of axioms would \emph{nearly} reproduce standard quantum mechanics, except that the expectation value of an operator \(\OO\) cannot in general be taken to be
	\(\bra\OO\ket\qu\bra\j|\OO|\j\ket\ ,\) but rather
	\be \bra\OO\ket=\sum_{\vec Q}|\bra\vec Q|\j\ket|^2\bra \vec Q|\OO|\vec Q\ket=\Tr(\rho\,\OO)\ ; \quad\rho=\sum_{\vec Q}\rho_{\vec Q}|\vec Q\ket\bra\vec Q|\ . \eel{densitym}
This means that we will not consider \emph{ontological} states to be in a quantum superposition, but we do allow density matrices. Note, that Eq.~\eqn{densitym} is not the most general density matrix used in quantum mchanics, but a subset. If, however, the states \(|\vec Q\ket\) are not known in a given quantum system, then it will also be difficult to tell whether a given density matrix can be cast in this form or not. The CA interpretation asserts that we can limit ourselves to density matrices that can be cast in this form.
 
Thus, a subtle departure from standard quantum mechanics is that all `physical', ontological states will be the \(|\vec Q\ket\) states themselves. Then it is also natural to assume that \emph{course grained observables}, such as the positions of cars and planets, but also the position of indicators on a detector, are representated by carefully chosen, statistical arrangements of the automaton states \(\vec Q_i\), and therefore the classical observables are diagonal in \(\vec Q_i\). This means that the ontological states \(|\vec Q_i\ket\) will automatically \emph{collapse} into macroscopically ovserved states, so that, within this interpretation schame, there is no measurement problem: collapse takes place totally automatically\cite{GtHcollapse}, while the states continue to obey the Schr\"odinger equation \eqn{schro}. Thus, we arrive at the following axiom:
	\bi{(vi)} All ``classical" states, such as planets, people, and indicators of detectors,  are CA states.  Therefore, at the end of an experiment, all observed quantities are described by operators that are diagonal in the ontological \(|\vec Q\ket\) basis.\ei

Apart from the claim that we can say something about the `physical', or `ontological' states, this theory is equivalent to standard quantum mechanics. Our mathematics allows us to consider superimposed states, but, as long as we stay in the ontological basis, these superpositions seem to be meaningless. The reason why we use superimposed states all the time in quantum mechanics is because we have not yet been able to identify its ontological basis; it may well require detailed descriptions of what goes on at the Planck scale, which are out of our reach today.

The above was motivated by our attempts to formulate the quantum rules for a theory of gravity that would allow for compact, finite universes to evolve.\cite{Carlip}\cite{Barrow} For such universes, it appears to make little sense to put them into quantum superpositions of states. Since the hamiltonian is conserved, finite systems that do not interact with the outside world must always be in eigen states of the hamiltonian, but this can also be interpreted as saying that the \emph{time coordinate} is a gauge parameter; the universe is invariant for shifts in the time parameter, and therefore, time as such is ill-defined and unobservable. Such finite systems, however, \emph{can} hop from one state into the next, a process that should be describable by distinct, unitary evolution operators, and the states that they act on could be interpreted as classical states, if treated in the correct basis. As will be explained, the most promising example of such a system can be transformed into superstrings, but we will come to that. First, one will have to address the no-go theorems.

\newsecl{Bell's inequalities}{Bell}
	J.S.~Bell's important and powerful observations\cite{Bell} have been reviewed so many times that another repetition of that is not necessary. Without going into details, we summarize his findings:\\
	A gedanken experiment can be performed, such as the Einstein-Podolsky-Rosen exp\-eriment\cite{epr} -- and there are numerous other variations on the same theme, often replacing continuous operators such as \(x\) and \(p\) by discrete and finite spin operators -- with two observers, who are looking at a set of entangled objects. Observer \(A\) (Alice) may make an observation allowing a filter \(\vec a\) and observer \(B\)(Bob) may use a filter \(\vec b\). \(A\) and \(B\) are allowed to modify the orientation of their filters immediately before they do their observation, \emph{at their free will}, and thus decide at the very last moment between various, mutually non-commuting operators for doing their observation. If, after the experiment, \(A\) and \(B\) compare their tables of observations, they will be unable to explain how classical signals can have travelled from the source of the particles towards  \(A\) and \(B\). These would obey inequalities, called the Bell inequalities, which are clearly not obeyed by the lists of data that they have in their hands.

An essential ingredient of this argument is that the actions of the source \(C\) of the entangled particles, cannot possibly depend on the later decisions made by both \(A\) and \(B\) to do their measurements. These decisions were made out of `free will'.

How exactly should we use this to invalidate the CA interpretation just formulated in the previous section? This is not so easy. First of all, there is only one set of ontological states nature is in, according to the CA interpretation, so one can never choose between two states that are not elements of the same basis. `Counterfactual' experiments are forbidden. So, working with eigen states of two non-commuting operators \(\vec a_1\) and \(\vec a_2\), as Alice wanted to do, and also the two settings \(\vec b_1\) and \(\vec b_2\) that Bob wanted to choose from, is not allowed in this formalism. In reality, we must assume that the states form a density matrix. Moreover, this density matrix is not allowed to have non-diagonal elements in the ontological basis. What makes this situation difficult to control is that these `ontological' states may have to be characterized at the Planck scale, or so, so that an actual description of what is happening is hard.

One must stick to following the gedanken experiment as closely as possible and see where it clashes with the cellular automaton idea. This happens right away when we consider the notion of `free will'. \emph{There is no free will in a cellular automaton.} Everything is pre-determined. This `superdeterminism', the demand that `counterfactual' gedanken experiments would be illegal, since only one real set of events, the events that actually take place, may be discussed, was already admitted to be an escape route by Bell himself.\cite{Bellcom}
``But", he adds, ``nobody wants superdeterminism". More precisely, even if, from a formal point of view, neither Alice not Bob can use their `free will', whatever they do use to make their decisions, should not affect the source \(C\) of the particles that they observe. The idea that atoms \( C\) would `know in advance' what decisions two people such as Alice  \((A)\)and Bob \((B)\) would later make, is considered to be \emph{ridiculous}. Signals would have to go from \(A\) and \(B\) backwards in time to reach \(C\) before the entangled particles are emitted.

Can there exist a form of `conspiracy' such that signals from the source and from Alice could reach Bob before he makes his decision? If that were the case, Bob's decision is actually far from free; it depends on what Alice does, what the source has done, and possibly, vice versa, Alice's decision depends on what Bob has been thinking as well. When phrazed this way, also this scenario sounds unlikely, but there is a better way of formulating what appears to be happening. 

Neither Alice nor Bob are acting out of free will, which means that their actions have their roots in the past. Consider the data of the system to be distributed over a Cauchy surface that evolves from past to future. At \(t=0\), the surface passes through the source \(C\) while it emits two entangled particles. At that same moment, other data on the Cauchy surface are configured in such a way that Alice will later make a decision \(\vec a\) and Bob will later make a decision \(\vec b\). Only if we change the data on this Cauchy surface, the later decisions \(\vec a\) and \(\vec b\) can at all be modified. Thus, even though they are entangled (possibly in a totally classical sense), we can characterize the data on the Cauchy surface as \(\vec a\), \(\vec b\) and \(C\). One can now derive that \(\vec a\), \(\vec b\) and \(C\) must be \emph{correlated}. the correlation function can be computed, and, in the case of a standard Bell experiment, the 3-point correlation function turns out to be non-trivial. In contrast, as soon as we average over all possible \(\vec a\), or just over all \(\vec b\), or the source \(C\), we get the 2 point correlators, and these functions can disappear completely: there will be no 2 point correlations between Alice's decision \(\vec a\) and Bob's \(\vec b\).

This 3 point correlation function will have to be non-vanishing even if \(\vec a\), \(\vec b\) and \(C\) are all spacelike separated. Now, our point is that this is not alarming at all. \emph{Correlation functions between observables, even if embedded in the vacuum state, and if they are spacelike separated, can be non-vanishing.} This has always been a central point in quantum field theory. In fact, one can derive all scattering amplitudes by analytical continuation of the spacelike correlators. For instance, even the 2 point correlators do not vanish: they are the propagators of the theory, whose poles in the complex plane betray the eigen values of the hamiltonian.

Thus, according to quantum field theories, the spacelike correlation functions not only do not vanish, they allow us to predict the future, in principle, since they give us the entire spectrum of the hamiltonian. Note that there is nothing spooky about these correlations; they certainly also occur in classical theories. Their existence simply means that variables that depend on phenomena in the intersection of their past light cones, may well be correlated for that reason.

This explanation is usually also dismissed. It is called a `conspiracy theory', and that is considered to be \emph{disgusting}. But are `disgusting', or `ridiculous',   valid arguments in a mathematical proof? We have reasons to doubt that. Rather than saying that there are `spooky signals' going around, we could also say that the laws of nature cause correlations, these correlations may even be controlled by various types of conservation laws. The behavior of our `hidden variables' may also be controlled by conservation laws. Perhaps these laws are not `disgusting'.

\newsecl{The CA as a universal computer}{CAuniversal}

Now the above is not at all the complete answer to Bell. In ordinary classical statistical systems, regardless whether they have stochastic elements built in or are completely deterministic, we usually do \emph{not} observe emergent quantum mechanical behavior. Where and when could apparent quantum features of a cellular automaton arise? Can something be classical and quantum at the same time?

What makes an automaton a \emph{cellular automaton}\cite{Fredkin}\cite{Toffoli}\cite{Feynman}\cite{Wolfram} is the additional requirement that the data are distributed over `cells'. These cells could be arranged in a \(D-1\) dimensional lattice, if \(D\) is the dimension of space-time, usually identified as \(D=4\). The evolution law is assumed to dictate how these cells evolve in time, and that the laws for each cell will only involve the data in its immediate neighbors and its own data. Signals are then limited to a maximal velocity, usually taken to be that of light.

In all other respects, the evolution law may be as complex as one likes, and the number of distinct data in each cell is not necessarily limited, although we usually do assume it to be some finite number in each cell.

As soon as one has enough freedom to choose the evolution law, it is not difficult to convince onself that the automaton will belong to a class that is called `universal'. \cite{universal} With this, we mean that it can, in principle, handle any calculational assignment, including the calculation as to what will happen in any \emph{other} cellular automaton. Thus, with `universal', we mean that the details of an automaton may become invisible: any automaton is as good as any other. Now, the present author suspects that this definition of `universality' is too crude. We could subsequently ask which specific evolution law of a cellular automaton would be the optimal one to describe our world? Quite likely there will be many solutions to this problem, but only a very small set will be efficient enough to reproduce exactly the physical world and nothing else.

Once such an evolution law has been identified, we can ask how it effectively gives rise to the physical phenomena that we observe. The presently observed or observable phenomena all take place at a scale that is some 16 orders of magnitude away from the Planck scale, which is quite possibly the scale of the CA. To cover these 16 orders of magnitude we have to make a scale transformation linking one world to the other. This is what in field theory is called the \emph{renormalization group}\cite{Callan}\cite{Symanzik}. In ordinary statistical physics, often these group transformations are extremely crude and considerable amounts of information is lost on the way. Only in the quantum field theoretical approach the action of these group transformations can be made much more precise, so that we can keep the notion of hilbert space in the process. Indeed, as anticipated in our axioms, the transformations do require unitary transformations of the basis of Fock space that totally deform Hilbert space once we crossed the 16 orders of magnitude. So, in our view, it should not be surprising that we have lost track completely of what the `ontological' basis of our Hilbert space should have been.

However, the notion of a universal computation device can now also be brought forward to argue that such computers cannot be guaranteed to yield Bell like inequalities when we avarage over correlated expressions. One cannot force the outcomes of universal computations to be free from superficially weird looking correlations.

All these considerations taken together were a motivation to proceed with the CA interpretation. It may be rewarding to study it more closely and worry about Bell's (important) inequalities later. A litmus test will be: \emph{Find a good example of a CA \(\leftrightarrow\) quantum mapping!}\cite{GtHCA},\cite{GtHbosonic}

\newsecl{The Superstring}{sustr}

Superstring theory\cite{GSW}\cite{Polchinski} is an extremely delicate mathematical construction that appears to serve well as a description of Planck length physics. It appears not only to unify bosonic gauge interactions, fermionic matter particles and scalar fields that can generate mass, it also generates quantum excitations that serve as gravitons, generating the gravitational field. This theory is often hauled as the most promising candidate for a realistic description of all existing forces and particles. However, its fundamental internal logic is difficult to fathom. ``This theory is even stranger than quantum mechanics", according to some of its proponents. Should we believe this? Gell-Mann had a more sensible remark: ``The world seems to become more and more complex, \emph{until you reach a new level of understanding}. Then things become simple again."\cite{GellMann}

We will now bring forward our reasons to believe that superstring theory is not only simpler than quantum mechanics, it is also simpler than classical mechanics. In classical mechanics, the predictive power of theories is obstructed by a fundamental feature called \emph{chaos}.\cite{chaos} This means that the physical data of a classical mechanical system require real numbers with infinite precision. Most real numbers in the data describing planets orbiting the Sun are known only up to a limited number of decimal places. But only if the entire, infinite sequences of decimal places are given, classical mechanics can be used as a deterministic theory. 

We now report our finding\cite{GtHsustr} that the superstring is mathematically equivalent to a deterministic cellular automaton that only processes integers, in discrete time steps. Thus the data form completely finite sets, and such theories should, in principle, allow for infinitely precise predictions. In practice, however, these data will be arranged in Planckian dimensions, and for that reason they will never be available to us. Our point is that such a theory will be on better foundations than even classical mechanics. The mathematics of this duality relation is straightforward and will be explained here. Physical aspects are still somewhat mysterious, not in the last place because we do not quite understand its `disgusting' conspiracy properties.

\newsecl{Mappings}{maps}

To understand our claims about the superstring, let us first phraze an important result:
\begin{quotation} \emph{\ \   \hskip-15pt There is a unitary transformation between:}
 \bi {\bal} Hilbert space spanned by the eigen states of a real number operator \(x\),\\[2pt]
\hbox{}\hskip-17pt \emph{\ and}
\itm {\bal} Hilbert space spanned by the eigen states of a pair \(\{Q,\,P\}\) of integer-valued operators. \ei
\end{quotation}
Indeed, in some of the more interesting cases, the integers may evolve classically while the real number operator can only evolve quantum mechanically. The most promising approach here is to try to extend Hamilton's beautiful canonical variables \(\vec q\) and \(\vec p\). In the procedure that will be explained now\cite{GtHPQ}, we replace them by two (sets of) integers, \(Q_i\) and \(P_i\).

Consider first a single set of integers \(Q\in{\mathbb Z}\). We assign an element of a basis of Hilbert space to each point in  \({\mathbb  Z}\). Next, we transform to a different basis. Consider a variable  \(\eta\) in the interval \((-\half,\half]\). Define, for ease of notation, a new basis for exponentials and logarithms:
	\be\eps=e^{2\pi}\approx 535.5\ . \eel{e2pi}
Then\fn{In different publications, we sometimes use different sign conventions here, but within a single paper such as this one, we try to keep our conventions consistent.}, we have the `\(Q\)-raising operator' \(\eps^{i\eta}\):
	\be e^{i\eta}|Q\ket=|Q+1\ket\ ; \qquad \bra\eta|Q\ket=\eps^{iQ\eta}\ ,\qquad  \bra Q|\eta\ket=\eps^{-iQ\eta}\eel{lowering}
(The advantage of the base \(\eps\) of Eq.~\eqn{e2pi} should be clear: these wave functions here are easy to normalize).
The operator \(\eta\) itself is found by using the discrete fourier transform on the unit interval:
	\be |\eta|<\half\ ,\quad \eta=\sum_{N=-\infty}^\infty \a_N\eps^{iN\eta}\ ,\quad \a_N=
		\int_{-\half}^\half\eta\eps^{-iN\eta}\dd\eta={i(-1)^N\over 2\pi N}\ \hbox{ if }\ N\ne 0\ ,   \eel{etafourier}
while \(\a_0=0\). This gives us the matrix elements of the operator \(\eta\) in the \(Q\)-basis:
	\be \bra Q_1|\eta|Q_2\ket={i\over 2\pi}(1-\d_{Q_1Q_2}){(-1)^{Q_1-Q_2}\over Q_1-Q_2} \eel {etaQelements}  (it vanishes if \(Q_1=Q_2\)). From this, we find the commutator:
	\be \bra Q_1|[\eta_Q,\,Q]|Q_2\ket &=&(Q_2-Q_1)\bra Q_1|\eta_Q|Q_2\ket\ =\nm\\
	{i\over 2\pi}\left(\d_{Q_1Q_2}-  (-1)^{Q_1-Q_2}\right)&=&\bra Q_1|\,{i\over 2\pi}\Big({\mathbb I}-|\j\ket\bra\j|\Big)\,|Q_2\ket\ ,\eel{etaQcomm}
where the \emph{edge state} \(|\j\ket\) is defined by \(\bra Q|\j\ket=(-1)^Q\).
Apart from this edge state, which we shall often ignore, we see that \(\eta\) and \(Q\) behave as a coordinate and its associated momentum operator.  This means that Hilbert space spanned by the integers can be easily mapped onto a Hilbert space spanned by the functions on a circle, or, if we ignore the edge states, functions on the interval \((-\half,\half]\) ; the edge states can be seen to be located on the seams: \(\bra\eta|\j\ket=\d(\eta-\half)\).
Functions that ar continuous on the circle are orthogonal to the edge state.

Consider two such integers, \(Q\) and \(P\).  Now, we can construct a single real-number operator from this pair. For a start, consider a real number \(q\in(-\infty,\infty)\). Then, let the integer \(Q\) be the closest integer to \(q\), or, \(Q=\hbox{round}(q)\), and \(\eta_P=q-Q\). This uniquely defines the Hilbert space spanned by the real numbers \(|q\ket\) as a product space of states \(|Q,\,\eta_P\ket=|Q\ket|\eta_P\ket\). If \(\eta_P\) is taken to be the `position operator' for the integer `momentum operator' \(P\), we find that the fourier transform now defines the states \(|q\ket\) as a unitary superposition of states \(|Q,\,P\ket\). Thus, we have 
% (a sign switch had to be made in the relation between \(P\) and \(\eta_P\))
	\be q\equiv Q+\eta_P\ ;\quad \bra Q,\eta_P|\j\ket\  = \sum_{P=-\infty}^\infty\eps^{iP\eta_P}\bra Q,P|\j\ket\iss\bra q|\j\ket\ . \eel{qPQ}
From this, we can also use the standard fourier transformation to transform to the real number variable \(p\), which in our notation obeys
	\be\bra q|p\ket=\eps^{ipq}\ ;\quad \bra p|\j\ket=\int_{-\infty}^\infty\eps^{-ipq}\dd q\bra q|\j\ket\ . \ee
If we write \(p=K+\k\), where \(K\in\mathbb Z\) and \(\k\in(-\half,\half]\), one finds
	\be\bra p|\j\ket=\bra K,\k|\j\ket=\sum_{Q,P}	{\sin\pi\k\over\pi}\ {(-1)^{K-P}\eps^{-i\k Q}\over K-P+\k}\ \bra Q,P|\j\ket\ .\eel{pPQ}
Now the kernel in this summation resembles the Kronecker delta somewhat, as it maximizes at \(P=K\), but there is a disturbing asymmetry between Eqs~\eqn{qPQ} and \eqn{pPQ}. \def\edge{{\mathrm{edge}}}

Also, the convergence when \(|P-K|\) becomes large is slow, which causes complications in later calculations. This is due to the edge states again. We found that we can improve the situation by removing most of the edge states. Only an edge state of the form \(\bra\eta_P,\eta_Q|\j\ket=\d(\eta_P-\half)\d(\eta_Q-\half)\) cannot be avoided. We found that further transformations lead to symmetric sets of matrix elements, which also converge faster. It was found that the real-valued operators \(q\) and \(p\) can be written as
	\be q=Q+a_Q\ ,&& p=P+a_P\ ;\nm\\
	\bra Q_1,\,P_1|a_Q|Q_2,\,P_2\ket\ =&&{(-1)^{P+Q+1}\ iP\over 2\pi(P^2+Q^2)}\ ;\crl{aQQP}
	\bra Q_1,\,P_1|a_P|Q_2,\,P_2\ket\ =&&{(-1)^{P+Q}\ iQ\over 2\pi(P^2+Q^2)}\ ,\eel{aPQP}
where \(P=P_2-P_1	\) and \(Q=Q_2-Q_1\).

We derive:
	\be [q,p]={i\over 2\pi}(\mathbb I-|\j_\edge\ket\bra\j_\edge|)\ ,\quad \bra Q,P|\j_\edge\ket=(-1)^{P+Q}\ . \eel{edgePQ}
This is as close to the canonical commutation rule as one can get. It means that, if we want to obtain continuum physics, we have to limit ourselves to states that are orthogonal to the single edge state \(|\j_\edge\ket\). Since \(|\j_\edge\ket\) stretches over the entire `universe' and over all \(P\) values, we think that this constraint has no importance for physics limited to some given region.

\newsecl{Quantum field theory in 1+1 dimensions}{1+1}
	The \((Q,P)\leftrightarrow(\hat q,\hat p)\) mapping is particularly significant for field theories in one space, one time dimension. Take the free quantum field theory, having field variables \(\f(x,t)\) and a canonical  momentum field \(p(x,t)\) obeying commutation rules
		\be [\f(x,t),\,p(x',t)]= {i\over 2\pi}\d(x-x')\ ,\qquad[\f(x,t),\,\f(x',t)]=[p(x,t),\,p(x',t)]=0\ . \eel{fieldcomm}
The Klein-Gordon equation,
	\be (\pa_x+\pa_t)(\pa_x-\pa_t)\f(x,t)=0\ , \eel{KG}
implies the existence of left-movers \(\f^L(x+t)\) and right-movers \(\f^R(x-t)\):
	\be\f(x,t)=\f^L(x+t)+\f^R(x-t)\ ,&& p(x,t)=\half a^L(x+t)+\half a^R(x-t)\ ,\crl{LRmovers}
	a^L(	x+t)=p(x,t)+\pa_x\f(x,t)\ , && a^R(x-t)=p(x,t)-\pa_x\f(x,t)\ . \eel{aLR}
In fact, up to some coefficients, these are fourier transforms of the familiar particle creation and annihilation operators.  We have the hamilton density
	\be \HH=\half(p^2+(\pa_x\f)^2)=\quart({a^L}^2+{a^R}^2)\ . \eel{fieldham}
In terms of the left-and right movers, the commutation rules are
	\be\hskip-10pt [a^L,\,a^R]=0\ , \quad[a^L(x),\,a^L(y)]={i\over\pi}\pa_x\d(x-y)\ ,\quad [a^R(x),\,a^R(y)]={-i\over\pi}\pa_x\d(x-y)\ . \eel{aLaRcomm}
Replacing the spacetime continuum by a (dense) lattice, one sees that these commutation rules can be replaced by
	\be \hskip-6pt [\f(x,t),\,p(y,t)]={i\over2\pi}\d_{x,y}\ ,\quad[a^L(x),a^L(y)=-[a^R(x),a^R(y)]=\pm{i\over 2\pi}\ \hbox{ if }\ y=x\pm 1\ . \eel{commlattice}

Now, suppose we introduce integer valued operators \(A^{L,R}(x)\) and their associated momentum operators \(\eta^{L,R}(x)\), obeying
	\be [\eta^L(x),\,A^L(y)]=[\eta^R(x),A^R(y)]={i\over 2\pi}\d_{x,y}\ , \quad[A^L,A^R]=[A^L,\eta^R]=0\ ,\ \hbox{etc.} \eel{etaAcomm}
(ignoring the usual edge states), one finds that we can write
	\be a^L(x)=A^L(x)+\eta^L(x+1)\ ,\qquad a^R(x)=A^R(x)+\eta^R(x-1)\ , \eel{aAeta}
so that the commutation rule \eqn{commlattice} is automatically obeyed.	

With some more advanced mathematics, one can reduce the effects of the edge states to a minimum (they cannot be iognored completely).
	
The importance of this procedure, and the reason why it \emph{only} works in one space, one time dimension, is that the time evolution of the \(a^{L,R}\) fields involves nothing more than shifts in \(x\) space, without further linear transformations such as additions or subtractions. Linear transformations in the variables \(A^{L,R}\) and \(\eta^{L,R}\) do not lead to similar operators where the \(A\) are integer and the \(\eta\) stay in the interval \((\-\half,\,\half]\).

We now observe that, in string theory, the \(\f\) fields are the left- and right moving coordinates \(X^\m\) in a \(D\) dimensional space-time. For convenience, we had chosen the world sheet lattice to have lattice length \(a=1\), but it is easy to verify that, if we had chosen any other lattice length, the relation between the quantized variables \(A(x)\) and the space-time coordinates \(X^\m(x)\) would remain the same. In other words, just because we wished to relate the commutation rules  \eqn{fieldcomm} and \eqn{aLaRcomm} to the commutation rules \eqn{etaAcomm}, and have the relations \eqn{aAeta} such that the real numbers are smoothly covered, we find that the classical system has its spacetime coordinates \(X^\m\) defined on a grid, and the grid length is fixed. Re-inserting the usual string units, one finds a space-time lattice with lattice length\cite{GtHsustr}
	\be a_{\hbox{spacetime}}=2\pi\sqrt{\a'}\ , \eel{spacetimealfa'}
a remarkable result. Note, that the classical theory has its string equations simply formulated in terms of the integer valued operators \(A^{L,R}(x,t)\) (where \(x\) and \(t\) are the world sheet variables usually denoted as \(\s\) and \(\t\). From the field equations \eqn{KG} -- \eqn{aLR}, one derives that the integer-valued string coordinates obey simple classical equations,
	\be X^\m(\s,\t+a)+X^\m(\s,\t-a)=X^\m(\s+a,\t)+X^\m(\s-a,\t)\ . \eel{discrstringeqs}
	
These equations must be assumed to describe the transverse string coordinates only. The longitudinal and timelike coordinates have to be derived from the usual string constraint equations. This part of the quantum theory remains unaffected. It means that, only in 26 dimensions, or in 10 dimensions if we add the fermionic degrees of freedom (see the next section), the quantum theory has an enhanced symmetry: Lorentz invariance.	

The string equation \eqn{discrstringeqs} is not quite as simple as it looks, since the world sheet coordinates \(\s\) and \(\t\) must still be taken to be on a world sheet lattice -- although the grid size of that can be taken to be arbitrarily small.

\newsecl{Fermions}{fermions}
	Usually, one assumes supersymmetry on the string world sheet. This means that one has a number of fermionic degrees of freedom there that matches the bosonic (coordinate) degrees of freedom. These are chiral fermions in 1+1 dimensions, also massless and moving with the speed of light. They obey Dirac equations,
	\be (\g_+\pa_-+\g_-\pa_+)\j=0\ , \eel{Dirac} 
where the Dirac matrices \(\g_\m\) are \(2\times 2\) matrices and the spinor has 2 components. From this, one finds
	\be\j	_A^\m(x,t)=\pmatrix{\j_L^\m(x+t)\cr\j_R^{\m\vphantom{|^|}}(x-t)}\ . \eel{spinorsol}
The corresponding classical theory has Boolean degrees of freedom 	\(s(x,t)=\pm 1\), obeying the equations
	\be s^\m(x,t+1)=s^\m(x-1,t)\,s^\m(x+1,t) s^\m(x,t-1)\ , \eel{Booleeq}
which split into left- and right movers:
	\be s^\m(x,t)=s^\m_L(x+t)\,s^\m_R(x-t)\ . \eel{BooleLR}	
	
To turn this system into a fermionic theory, all we have to do is perform a Jordan-Wigner transformation.\cite{JW} Further details are exposed in Ref.\cite{GtHsustr}.

Thus, superstrings are distinguished from bosonic strings in that, the classical system not only has discrete coordinates \(X^\m(\s,t)\), but also 	Boolean variables \(s^\m(\s,\t)=\pm 1\).

So-far, we only handled strings with infinite lengths. To obtain finite length closed strings, one has to introduce periodic boundary conditions. This has not yet been done in detail. Also, the fermionic system should include constraint equations that produce \(\j^\pm_A\), the longitudinal components of the fermionic fields. Here again, one should keep the quantum formalism of the superstring unchanged.

\newsecl{String interactions}{int}

The above would only have been useful to describe freely moving strings, whose basic excitations would correspond to free particles. One might have been worried that inserting interactions would be impossible in a classical theory if we wanted that to be mapped on a quantum system. This, however, does not seem to be true. We can introduce deterministic interactions at the classical level. The basic interaction we are thinking of is the \emph{exchange interaction}, see Fig.~\ref{stringint.fig}. If two strings hit the same spacetime point \(X^\m\), two arms are exchanged.
		
\begin{figure}[h] \setcounter{figure}{0} \includegraphics[width=100mm]{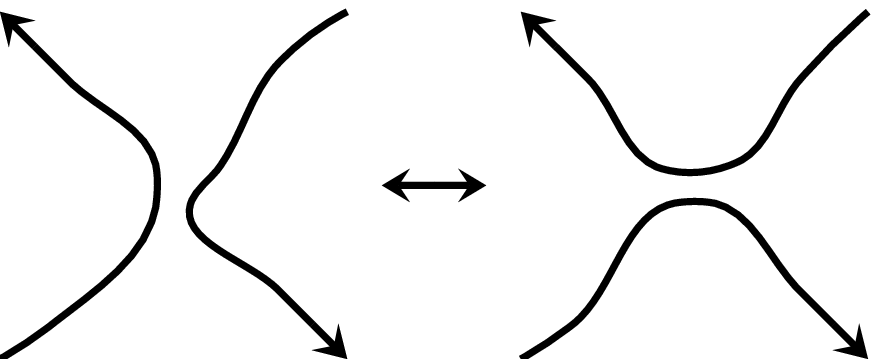} \begin{quotation} \begin{caption}
	{ Basic string interaction by exchanging arms. This is only unambiguous if the strings are also oriented (arrow). 
	}\label{stringint.fig}	\end{caption} \end{quotation}
\end{figure}

This interaction should also generate closed, interacting,  oriented  strings.	
If we want this to be a deterministic theory, the exchange should occur with probability 0 if the strings do not hit the same point in \(D\) space, and probability 1 if they do. Thus, we must conclude that the string constant  \(g_s\) is fixed to \(g_s=1\),  and the strings must be \emph{oriented}.

Ambiguity does arise if we compactify the surplus dimensions, which can be done in mathematically distinct ways. 
This should be carefully investigated.

\newsecl{Conclusions}{conc}

A number of simple, totally classical physical models admit a dual mapping to models that are quantum mechanical. Most of these appear to be physically fairly trivial, with the exception of the cellular automaton; these however seem  to be dual to quantum models that do not resemble the real world very much. Also, a generic cellular automaton model would not obviously have any Lorentz invariance, or even galilean invariance: a `particle' at rest cannot easily be turned into a particle moving with any arbitrary velocity.

A very interesting exceptional case seems to be an automaton that is dual to strings or even superstrings. Superstring theory is suspected to allow for realizations that resemble the real world fairly closely; indeed it has been conjectured that the world we live in \emph{is} a superstring world. We showed explicitly how the bulk equations of this sytem are dual to a classical model, where physical space-time is a lattice with a precisely defined lattice constant (being proportional to \(\sqrt{\a'}\), where \(\a'\) is the string slope parameter, usually assumed to be fairly close to the Planck length, but a bit larger).

As for the string's periodic boundary conditions and the dynamics of its interactions, we have as yet only made conjectures; it is not clear whether these details indeed accurately describe 	a desired theory or model.

We find the existence of these mappings surprising, since, usually, they are considered to be impossible. The ensuing quantum models must violate Bell's important inequalities, or, more generally, must allow for the generation of quantum mechanically entangled particle states. Although we claim that there seems to be no compelling reason for these models to forbid quantum entanglement -- as we can envision any quantum state we like -- this still raises the question how information is transported in any of the typical gedanken experiments that have been proposed -- and in fact carried out.\cite{Aspect}

We suspect that the answer to such questions will resemble what is called `conspiracy' in the literature: the notion that classical correlation functions seem to `conspire' to generate entangled systems. If such conspiracy can be linked to conservation laws for the hidden variables (which here are simply the physical orientations of the `ontological' basis elements, then `conspiracy' may become easier to understand and accept as a natural feature of nature after all.

\end{document}
	
	N.H.  Margolus,  Physica \textbf{10D}  (1984)  81;